\documentstyle[12pt]{article}
\makeatletter
\@addtoreset{equation}{section}

\makeatother

\begin{document}

\begin{flushright}
Jul 2001

KEK-TH-778
\end{flushright}

\begin{center}

\vspace{5cm}

{\Large Melvin Background in Heterotic Theories} 

\vspace{2cm}

Takao Suyama \footnote{e-mail address : tsuyama@post.kek.jp}

\vspace{1cm}

{\it Theory Group, KEK}

{\it Tsukuba, Ibaraki 305-0801, Japan}

\vspace{4cm}

{\bf Abstract} 

\end{center}

We give an interpretation of the non-supersymmetric heterotic theories as the supersymmetric 
heterotic theories on Melvin background with Wilson lines. 
The coincidence of the partition functions is shown for most of the non-supersymmetric 
theories. 
We also discuss tachyonic instabilities in terms of non-trivial background fields. 
The behavior in the strong coupling region is investigated by using the duality relations 
of the supersymmetric theories. 

\newpage

\vspace{1cm}

\section{Introduction}

\vspace{5mm}

To understand the structure of string theory, one has to know its vacuum state. 
Although there are infinitely many perturbative vacua, some of which 
resemble our world, little is known what the nonperturbative vacuum is. 
To analyze the vacuum we need more sophisticated 
understanding of the nonperturbative dynamics of the theory, and this problem seems to be 
still beyond our reach. 
One possible approach to this problem would be to construct the theory nonperturbatively 
\cite{Matrix}. 

It is necessary to understand the (in)stabilities of the perturbative vacua. 
Recent reseach originated by Sen \cite{Sen} have clarified the stabilization mechanism of a 
perturbative instability associated with open string tachyons. 
This is closely related to the instability of D-branes, and the stabilization 
corresponds to the decay of the D-branes. 
The open string field theory \cite{WittenSFT} plays a crucial role in the investigations of 
the tachyon condensation \cite{OSFT}. 

There is another perturbative instability due to closed string tachyons. 
An interesting conjecture was made in \cite{CostaGutperle}\cite{fluxbrane} 
(see also \cite{RT}), suggesting that 
the endpoint of the closed string tachyon condensation in Type 0A theory 
can be interpreted as a supersymmetric vacuum. 
Type 0A theory is interpreted as Type IIA theory on the Melvin background \cite{Melvin} which 
breaks all the supersymmetries. 
And it is also shown that the tachyonic instability in Type 0A theory can be related to the 
instability due to pair creations of D-branes. 
The above results might indicate that Type 0A theory decays into Type IIA theory after the 
closed string tachyon condensation. 

In our previous paper \cite{Suyama}, we tried to apply a similar interpretation to the 
non-supersymmetric heterotic theories \cite{nonSUSYhet}\cite{nonSUSYhet2}\cite{list}, and 
some of them can be understood as the supersymmetric theories on the Melvin background. 
This would suggest that such understanding of the closed string tachyons is a general 
one, not specific to Type 0A theory. 

In this paper, we extend our investigations to all of the non-supersymmetric heterotic 
theories in ten dimensions. 
Although there is one exception, the other theories can be related to the supersymmetric 
heterotic theories. 
So this result leads us to a conjecture which suggests that the non-supersymmetric heterotic 
theories would decay into the supersymmetric ones. 
We also argue the instability due to the closed string tachyons in terms of the background 
fields in the supersymmetric theories. 

Since the non-supersymmetric theories can be related to the supersymmetric ones, one can 
discuss the strong coupling behavior of the former by using the duality relations of the 
latter. 
This enables us to see the validity of the decay picture beyond the perturbation theory. 

This paper is organized as follows. 
We briefly review the properties of the Melvin background (section 2) and the correspondence 
between Type 0A theory and Type IIA theory stated above (section 3). 
In section 4, we recall some facts about the non-supersymmetric heterotic theories. 
They are related to the supersymmetric theories in section 5 by showing the coincidence of 
the partition functions. 
The tachyonic instability is interpreted in terms of the background fields in the 
supersymmetric theories in section 6. 
In section 7, we discuss the strong coupling behavior of the non-supersymmetric theories. 
Section 8 is devoted to discussions. 
The stabilization of the heterotic $E_8$ theory, the exceptional case, is also argued.  
The explicit expressions of various partition functions discussed in section 5 are 
summarized in the appendix.

\vspace{1cm}

\section{Kaluza-Klein Melvin background}

\vspace{5mm}

Consider the following metric in M-theory \cite{CostaGutperle} 
\begin{equation}
ds_{11}^2 = \eta_{\alpha\beta}dx^\alpha dx^\beta+dr^2+r^2(d\theta+qdy)^2+dy^2.
  \label{metric}
\end{equation}
The indices $\alpha,\beta$ run from $0$ to $7$, and $q$ is a real parameter. 
We have used the polar coordinates $r,\ \theta$ in the 8-9 plane. 
The $y$-direction is compactified on $S^1$ with the radius $R$. 

By introducing the new coordinate $\tilde{\theta} = \theta+qy$, the metric (\ref{metric}) 
becomes the flat one, although the periodicity of $\tilde{\theta}$ is nontrivial, 
\begin{eqnarray}
&& y \sim y+2\pi mR, \\
&& \tilde{\theta} \sim \tilde{\theta}+2\pi n+2\pi mqR, 
  \label{period}
\end{eqnarray}
where $m,n$ are integers. 
Since the metric (\ref{metric}) is locally flat, this is a solution of the Einstein equation. 
This would still be a solution even if higher derivative terms are included, so this 
background is expected to be an exact solution of M-theory. 

The metric (\ref{metric}) describes a nontrivial spacetime in view of the lower dimensions. 
The dimensional reduction along the $S^1$ produces the following ten-dimensional (Type IIA) 
background in the string frame 
\begin{eqnarray}
ds_{10}^2 &=& \sqrt{1+q^2r^2}\ (\eta_{\alpha\beta}dx^\alpha dx^\beta+dr^2)
             +\frac{r^2}{\sqrt{1+q^2r^2}}d\theta^2, \\
e^{\frac43\phi} &=& 1+q^2r^2, \\
A_\theta &=& \frac{qr^2}{1+q^2r^2}.
  \label{IIA}
\end{eqnarray}
The dilaton diverges as $r\to\infty$, so this background cannot be analyzed by the 
perturbation theory. 
In addition, from (\ref{IIA}) one can see that there is a nonzero component of the R-R field 
strength
\begin{equation}
F_{89} = \frac{2q}{(1+q^2r^2)^2},
\end{equation}
which makes the perturbative analysis harder. 
The background (\ref{IIA}) describes a localized R-R flux around $r=0$ with 
the total flux $1/q$. 
This R-R flux configuration is called a fluxbrane or a F7-brane 
\cite{CostaGutperle}\cite{fluxbrane}. 
Further investigations on the fluxbranes and related issues can be found in 
\cite{fluxbrane2}\cite{motl}\cite{fluxbrane3}\cite{fluxbrane4}. 

\vspace{2mm}

From the identification (\ref{period}), one can see the periodicity in $q$. 
If $q=k/R$ for an integer $k$, $\tilde{\theta}$ has the period $2\pi$ and the spacetime 
is nothing but a direct product ${\bf R}^{10}\times S^1$, which is the same as that with 
$q=0$. 
Thus the parameter $q$ has the period $1/R$. 
This periodicity cannot be seen from the ten-dimensional background. 

When there exist spacetime fermions, the situation is more complicated. 
Suppose the case $q=1/R$. 
For the fermions to exist, the boundary condition along the $S^1$ should be periodic. 
Now the parallel transport of the fermions along the $S^1$ is accompanied by the $2\pi$ 
shift of $\tilde{\theta}$, so this gives an extra sign. 
In other words, although the metric is the same, the spin structure is different between 
$q=0$ and $q=1/R$. 
Therefore the period of $q$ should be $2/R$, when the spin structure is taken into account. 

\vspace{2mm}

The Melvin background is unstable. 
The instability is investigated in \cite{instability}. 
It is due to pair creations of 
D6-branes in the Type IIA picture, which is a similar phenomenon to the Schwinger pair 
creations in an electric field. 
It is shown in \cite{instability}\cite{CostaGutperle} that the decay mode is described by 
the Myers-Perry Kerr instanton \cite{instanton}, and the decay rate is zero only at $q=0$. 
So this result indicates that the Melvin background decays into the Kaluza-Klein background 
without R-R flux. 
On the other hand, as mentioned above, the Melvin background with $q=1/R$ (we will 
call it ``critical'' Melvin) can be interpreted as the Kaluza-Klein background with the 
twisted boundary condition for fermions along the $S^1$. 
In this picture, the instability of this background is described by an expanding bubble 
\cite{bubble}, corresponding to the Schwarzchild instanton. 
It is shown in \cite{bubble} that the Kaluza-Klein background is unstable against the 
nucleation of such bubbles, unless the spin structure is suitably chosen. 
If one chooses the periodic boundary condition, which corresponds to a supersymmetric 
background of string theory and M-theory, the decay through the bubble nucleation is 
forbidden by the topological reason. 
On the other hand, the critical Melvin has the different spin structure, so there is no 
reason to prevent it from the decay.

\vspace{1cm}

\section{IIA-0A duality}

\vspace{5mm}

It is discussed in \cite{Type0} that M-theory compactified on the twisted $S^1$ is 
equivalent to Type 0A theory, just like the ordinary $S^1$ compactification is equivalent to 
Type IIA theory. 
The evidence of this equivalence is the following. 

Compactify the background (\ref{metric}) further on $S^1$, say, along the 6-direction, and 
take the radius of the $S^1$ to be zero. 
Then the resulting ten-dimensional background is the Kaluza-Klein Melvin background without 
R-R flux, and the dilaton is a constant. 
So this background can be analyzed by the perturbative Type IIA theory. 

The one-loop partition function can be calculated exactly \cite{RussoTseytlin} 
\begin{eqnarray}
Z_q &=& \int \frac{d^2\tau}{\tau_2}\tau_2^{-4}|\eta(\tau)|^{-18}
        \sum_{m,n\in{\bf Z}}\exp\left[-\frac{\pi R^2}{\alpha'\tau_2}|n+m\tau|^2\right]
        \nonumber \\
    & & \hspace*{5mm}\times
        \left|\vartheta{1/2+mqR \brack 1/2+nqR}(0,\tau)\right|^{-2}
        \left|\vartheta{1/2+mqR/2 \brack 1/2+nqR/2}(0,\tau)\right|^8.
  \label{IIAPF}
\end{eqnarray}
We have used the theta function with characteristics
\begin{equation}
\vartheta{a \brack b}(\nu,\tau) 
   = \sum_{n\in{\bf Z}}\exp\left[\pi i(n+a)^2\tau+2\pi i(n+a)(\nu+b)\right].
\end{equation}
From the expression (\ref{IIAPF}) one can see the periodicity in $q$ as the periodicity of 
the theta functions. 
The period is $2/R$ since Type IIA theory contains spacetime fermions. 
At $q=1/R$ one finds
\begin{equation}
Z_{1/R} \sim \int \frac{d^2\tau}{\tau_2}\tau_2^{-5}|\eta(\tau)|^{-16}
             \left(|Z^0_0(\tau)|^8+|Z^0_1(\tau)|^8+|Z^1_0(\tau)|^8\right),
\end{equation}
if $R$ is sufficiently small. 
The RHS coincides with the partition function for Type 0A theory compactified on $S^1$ with 
vanishing radius. 
Thus we find that both Type IIA theory on the critical Melvin and Type 0A theory have the 
same mass spectrum in a certain limit. 

Suppose that this correspondence holds even before the dimensional reduction along the 
6-direction. 
Then the above result suggests that M-theory on the twisted $S^1$, which is equivalent to 
M-theory on the critical Melvin, is equivalent to Type 0A theory. 

\vspace{2mm}

According to the above interpretation of Type 0A theory, it is expected that the instability 
of the Melvin background (or of the Kaluza-Klein background with anti-periodic boundary 
condition) would 
correspond to the one due to the closed string tachyon in Type 0A theory 
\cite{CostaGutperle}\cite{fluxbrane}. 
The stabilization of closed string tachyons seems to be a more difficult problem to deal 
with than that of open string tachyons \cite{Sen}. 
So the above-mentioned interpretation would give us an insight into the closed string 
tachyon condensation.

\vspace{1cm}

\section{Heterotic theories in ten dimensions}

\vspace{5mm}

Type 0A theory discussed in the previous section can be regarded as a Type IIA theory 
twisted by a discrete goup ${\bf Z}_2$. 
This ${\bf Z}_2$ is generated by $(-1)^{F_s}$ where $F_s$ is the spacetime fermion number. 
Such a procedure enables us to construct a new string theory from the known one. 
This was investigated for the heterotic theories in \cite{nonSUSYhet}\cite{nonSUSYhet2}, and 
several ten-dimensional heterotic theories were discovered. 
And finally, the list of the heterotic theories in ten dimensions was completed in 
\cite{list}. 
There are nine theories, two of which are the well-known supersymmetric theories. 
Among the remaining seven, which have no supersymmetry, only one theory is 
tachyon-free, while the others contain tachyons in their mass spectra. 
The non-supersymmetric theories can be characterized by their gauge groups as follows. 

\vspace{1mm}

1. The $SO(16)\times SO(16)$ theory without any tachyon. 

\vspace{1mm}

2. The $SO(32)$ theory with tachyons in the ${\bf 32}$ representation of the gauge group. 

\vspace{1mm}

3. The $E_8\times SO(16)$ theory with tachyons in the $({\bf 1},{\bf 16})$. 

\vspace{1mm}

4. The $SO(8)\times SO(24)$ theory with tachyons in the $({\bf 8_s},{\bf 1})$. 

\vspace{1mm}

5. The $E_7\times SU(2)\times E_7\times SU(2)$ theory with tachyons 
  in the $({\bf 1},{\bf 2},{\bf 1},{\bf 2})$. 

\vspace{1mm}

6. The $U(16)$ theory with two tachyons in the singlet. 

\vspace{1mm}

7. The $E_8$ theory with only one tachyon. 

\vspace{2mm}

As shown in \cite{nonSUSYhet}, each theory, except for the $E_8$ theory, can be constructed 
from the supersymmetric theories 
by a ${\bf Z}_2$ twist which is generated by $(-1)^{F_s}\gamma_\delta$, where 
$\gamma_\delta$ is a shift operator in the internal momentum space, in view of the bosonic 
construction of the current algebra, with the shift vector $\delta$. 
Of course, the $\gamma_\delta$ satisfies $(\gamma_\delta)^2=1$. 

Consider an example; $\delta=(1,0^{15})$. 
It is convenient to discuss the internal left-moving degrees of freedom in the fermionic 
representation with fermions $\lambda_L^I$ $(I=1,\cdots,32)$. 
Then the action of the $\gamma_\delta$ is 
\begin{equation}
\gamma_\delta = e^{2\pi iJ_{12}},
\end{equation}
where $J_{12}$ is the generator of the $SO(32)$ which acts only on $\lambda_L^{1,2}$. 
This acts trivially on the vector representation of the $SO(32)$ and gives $-1$ for the 
spinor representation. 
Thus the untwisted sectors are
\begin{equation}
(left,\ right) = (NS+,\ NS+)\oplus(R+,\ R+).
\end{equation}
In the twisted sectors, the $\lambda_L^{1,2}$ obey the following boundary conditions 
\begin{equation}
\lambda_L^{1,2}(\sigma+2\pi) = e^{2\pi i}\lambda_L^{1,2}(\sigma).
\end{equation}
So this does not seem to change anything. 
However, the fermion number of the vacuum changes by the spectral flow. 
Therefore, in the twisted sectors the spectrum is restricted by the opposite GSO 
projection. 
The GSO projection of the right-moving sectors is also reversed and we obtain 
\begin{equation}
(NS-,\ NS-)\oplus(R-,\ R-). 
\end{equation}
The above spectrum coincides with that of the non-supersymmetric $SO(32)$ theory. 
The construction of other theories will be reviewed briefly in the appendix.

\vspace{1cm}

\section{Heterotic theories on Melvin background}

\vspace{5mm}

In this section, we will calculate partition functions of the supersymmetric heterotic 
theories on the Melvin background with various patterns of Wilson lines. 
It can be shown that the partition functions coincide with those of the non-supersymmetric 
heterotic theories reviewed in section 4, in a certain limit. 
This fact would be a piece of evidence for the equivalence between the non-supersymmetric 
heterotic theories and the supersymmetric heterotic theories on the Melvin background.

\vspace{5mm}

\subsection{Partition functions}

\vspace{3mm}

We consider the supersymmetric heterotic theories on the Melvin background with a general 
Wilson line. 
The worldsheet action is as follows,
\begin{eqnarray}
S &=& \frac1{4\pi\alpha'}\int d^2\sigma\left\{ \eta_{\mu\nu}\partial_aX^\mu\partial_aX^\nu
       +|\partial_aX+iq\partial_aYX|^2+(\partial_aY)^2\right\} 
      \nonumber \\
  & & \hspace*{-5mm}
      +\frac i{\pi}\int d^2\sigma\ \bar{S}_R^r\left(\partial_++\frac i2q\partial_+Y\right)
        S_R^r
      +\frac i{\pi}\int d^2\sigma\ \sum_{I=1}^{16}
        \bar{\lambda}_L^I\left(\partial_--iA_Y^I\partial_-Y\right)\lambda_L^I,
  \label{HetAction}
\end{eqnarray}
where $\mu,\nu=0,\cdots,6$ and $r=1,\cdots,4$. 
We have employed the Green-Schwarz formalism for the right-moving fermions $S_R^r$, while 
the left-moving fermions $\lambda_L^I$ are the RNS fermions. 
Although this action has nontrivial couplings between $Y$ and other fields, the theory is 
conformally invariant for any value of $q$ \cite{conformal}. 

An interesting property of the action (\ref{HetAction}) is that this can be reduced to a 
free action \cite{RussoTseytlin}. 
This is accomplished by the following field redefinitions
\begin{eqnarray}
&& X = e^{-iqY}\tilde{X}, \nonumber \\
&& S_R^r = e^{-\frac i2qY}\tilde{S}_R^r ,
  \label{redefine} \\
&& \lambda_L^I = e^{iA_Y^IY}\tilde{\lambda}_L^I. \nonumber 
\end{eqnarray}
Note that the Jacobian of these field redefinitions is trivial. 
The existence of the nontrivial background is now encoded in the boundary conditions for 
the new fields $\tilde{X},\ \tilde{S}_R^r,\ \tilde{\lambda}_L^I$. 

\vspace{2mm}

We will calculate the partition function on the torus with the modulus 
$\tau=\tau_1+i\tau_2$. 
To do this, we have to specify the boundary conditions. 
$X^\mu$ and $Y$ have the ordinary periodicity, 
\begin{eqnarray}
&& X^\mu(\sigma^1+2\pi,\sigma^2) = X^\mu(\sigma^1+2\pi\tau_1,\sigma^2+2\pi\tau_2)
                                 = X^\mu(\sigma^1,\sigma^2),  \\
&& Y(\sigma^1+2\pi,\sigma^2) = Y(\sigma^1,\sigma^2)-2\pi mR, \\
&& Y(\sigma^1+2\pi\tau_1,\sigma^2+2\pi\tau_2) = Y(\sigma^1,\sigma^2)+2\pi nR,
\end{eqnarray}
where $m,n$ are integers. 
The boundary conditions for the other fields defined in (\ref{redefine}) are twisted as 
follows
\begin{eqnarray}
&& \tilde{X}(\sigma^1+2\pi,\sigma^2) = e^{-2\pi imqR}\tilde{X}(\sigma^1,\sigma^2), \\
&& \tilde{X}(\sigma^1+2\pi\tau_1,\sigma^2+2\pi\tau_2) 
       = e^{2\pi inqR}\tilde{X}(\sigma^1,\sigma^2),  \\
&& \tilde{S}_R^r(\sigma^1+2\pi,\sigma^2) = e^{-\pi imqR}\tilde{S}_R^r(\sigma^1,\sigma^2), \\
&& \tilde{S}_R^r(\sigma^1+2\pi\tau_1,\sigma^2+2\pi\tau_2) 
       = e^{\pi inqR}\tilde{S}_R^r(\sigma^1,\sigma^2),  \\
&& \tilde{\lambda}_L^I(\sigma^1+2\pi,\sigma^2) 
       = \pm e^{2\pi imA_Y^IR}\tilde{\lambda}_L^I(\sigma^1,\sigma^2), 
  \label{leftBC1} \\
&& \tilde{\lambda}_L^I(\sigma^1+2\pi\tau_1,\sigma^2+2\pi\tau_2) 
       = \pm e^{-2\pi inA_Y^IR}\tilde{\lambda}_L^I(\sigma^1,\sigma^2) .
  \label{leftBC2} 
\end{eqnarray}
The plus (minus) signs in (\ref{leftBC1})(\ref{leftBC2}) are the ordinary ones for the R 
(NS) sector, respectively. 

The partition function for a left-moving fermion with a twisted boundary condition is easily 
calculated, for example, by using the operator formalism (see e.g. \cite{CMP}). 
For the fermion with boundary conditions
\begin{eqnarray}
&& \psi(\sigma^1+2\pi,\sigma^2) = -e^{2\pi i\theta}\psi(\sigma^1,\sigma^2), \\
&& \psi(\sigma^1+2\pi\tau_1,\sigma^2+2\pi\tau_2) 
       = -e^{-2\pi i\phi}\psi(\sigma^1,\sigma^2),
\end{eqnarray}
the partition function is, up to an overall factor, 
\begin{eqnarray}
Z^{2\theta}_{2\phi}(\tau) &\equiv& \vartheta{\theta \brack \phi}(0,\tau)\ \eta(\tau)^{-1} 
  \nonumber \\
&=& e^{2\pi i\theta\phi}q^{\frac{\theta^2}2-\frac1{24}}\prod_{n=1}^\infty
     \left(1+q^{n+\theta-\frac12}e^{2\pi i\phi}\right)
     \left(1+q^{n-\theta-\frac12}e^{-2\pi i\phi}\right),
\end{eqnarray}
where $q=e^{2\pi i\tau}$. 
The case of a twisted boson is similar. 
If $\psi$ is a bosonic field with the same boundary conditions, its partition function is 
$|Z^{2\theta}_{2\phi}(\tau)|^{-2}$. 

For the heterotic theories, relative phases of the partition functions for various 
sectors are relevant. 
These phases have been fixed in \cite{toroidal} by an anomaly argument. 
The resulting partition functions for the left-moving fermions are the followings,
\begin{eqnarray}
&& \mbox{(AA)}\hspace{5mm} e^{-\pi imn(a^I)^2}Z^{2ma^I}_{2na^I}(\tau), \hspace{1cm}
   \mbox{(AP)}\hspace{5mm} e^{-\pi ima^I(na^I+1)}Z^{2ma^I}_{1+2na^I}(\tau), \\
&& \mbox{(PA)}\hspace{5mm} e^{-\pi imn(a^I)^2}Z^{1+2ma^I}_{2na^I}(\tau), \hspace{7mm}
   \mbox{(PP)}\hspace{5mm} -ie^{-\pi ima^I(na^I+1)}Z^{1+2ma^I}_{1+2na^I}(\tau). 
\end{eqnarray}
Here (AA) etc. indicate the spin structures on the torus, and we have defined $a^I=A_Y^IR$. 

From the above results, one can construct the desired partition function as follows. 
\begin{eqnarray}
Z_q(A_Y^I) &=& \int\frac{d^2\tau}{\tau_2}\tau_2^{-4}|\eta(\tau)|^{-12}
   \sum_{m,n\in{\bf Z}}\exp\left[-\frac{\pi R^2}{\alpha'\tau_2}|n+m\tau|^2\right]
   |Z^{1+2mqR}_{1+2nqR}(\tau)|^{-2} \nonumber \\
& &   \exp\left[\pi imn\{(qR)^2-\sum_{I=1}^{16}(a^I)^2\}\right]e^{2\pi imqR}
      Z_L^{(m,n)}(\tau){Z^{1+mqR}_{1+nqR}(\tau)^*}^4
  \label{fullPF}
\end{eqnarray}
The contribution $Z_L^{(m,n)}(\tau)$ from the left-movers depends on which theory we 
consider. 
For the $SO(32)$ theory, 
\begin{eqnarray}
Z_L^{(m,n)}(\tau) &=& \prod_{I=1}^{16}Z^{2ma^I}_{2na^I}(\tau)
             +\prod_{I=1}^{16}e^{-\pi ima^I}Z^{2ma^I}_{1+2na^I}(\tau)
    \nonumber \\
                  & &+\prod_{I=1}^{16}Z^{1+2ma^I}_{2na^I}(\tau)
             +\prod_{I=1}^{16}e^{-\pi ima^I}Z^{1+2ma^I}_{1+2na^I}(\tau),
\end{eqnarray}
while for the $E_8\times E_8$ theory, 
\begin{eqnarray}
& & Z_L^{(m,n)}(\tau) \nonumber \\
&=& \left(\prod_{I=1}^{8}Z^{2ma^I}_{2na^I}(\tau)
             +\prod_{I=1}^{8}e^{-\pi ima^I}Z^{2ma^I}_{1+2na^I}(\tau)
             +\prod_{I=1}^{8}Z^{1+2ma^I}_{2na^I}(\tau)
             +\prod_{I=1}^{8}e^{-\pi ima^I}Z^{1+2ma^I}_{1+2na^I}(\tau)\right)
    \nonumber \\
& & \times\left(\prod_{I=9}^{16}Z^{2ma^I}_{2na^I}(\tau)
             +\prod_{I=9}^{16}e^{-\pi ima^I}Z^{2ma^I}_{1+2na^I}(\tau)
             +\prod_{I=9}^{16}Z^{1+2ma^I}_{2na^I}(\tau)
             +\prod_{I=9}^{16}e^{-\pi ima^I}Z^{1+2ma^I}_{1+2na^I}(\tau)\right). \nonumber \\
\end{eqnarray}

\vspace{5mm}

\subsection{Critical Melvin}

\vspace{3mm}

We are interested in the case with a special value of $q$, i.e. $q=1/R$. 
In this case, although the boundary condition for spacetime fermions on the $S^1$ is 
twisted, the background geometry is equivalent to the direct product ${\bf R}^9\times S^1$. 
Therefore it is natural to expect that a supersymmetric heterotic theory on the critical 
Melvin is equivalent to a non-supersymmetric heterotic theory, as in the IIA-0A case. 

To relate the partition function (\ref{fullPF}) to those of the non-supersymmetric theories, 
we have to choose the appropriate Wilson lines. 
These are summarized in the appendix. 
By the choice of the Wilson lines, the function $Z_L^{(m,n)}(\tau)$ has the following 
periodicity in $m$ and $n$,
\begin{equation}
Z_L^{(m+2,n)}(\tau) = Z_L^{(m,n+2)}(\tau) = Z_L^{(m,n)}(\tau).
  \label{periodicity}
\end{equation}
This means that $Z_L^{(m,n)}(\tau)$ depends only on $m,n\ \mbox{mod}\ 2$. 
Thus, for example, $Z_L^{(2m,2n)}(\tau)$ is the partition function for untwisted fermions 
and $Z_L^{(2m+1,2n)}(\tau)$ is the one for fermions twisted in the $\sigma^1$ direction. 
Another property of the Wilson lines is 
\begin{equation}
\sum_{I=1}^{16}(a^I)^2 = \mbox{integer}.
  \label{sum}
\end{equation}

Consider first the case in which the sum (\ref{sum}) is an odd integer. 
Under these conditions, the full partition function (\ref{fullPF}) can be rewritten as 
follows, 
\begin{eqnarray}
Z_{1/R}(A_Y^I) 
= \int \frac{d^2\tau}{\tau_2}\tau_2^{-5}|\eta(\tau)|^{-16}\left\{ 
    \sum_{k,l\in{\bf Z}}\exp\left[-\frac{\pi R^2}{\alpha'\tau_2}|(2l)+(2k)\tau|^2\right]
    Z_L^{(0,0)}(\tau){Z^{1+2k}_{1+2l}(\tau)^*}^4 \right. && \nonumber \\
   +\sum_{k,l\in{\bf Z}}\exp\left[-\frac{\pi R^2}{\alpha'\tau_2}|(2l+1)+(2k)\tau|^2\right]
    Z_L^{(0,1)}(\tau){Z^{1+2k}_{2l}(\tau)^*}^4 && \nonumber \\
   +\sum_{k,l\in{\bf Z}}\exp\left[-\frac{\pi R^2}{\alpha'\tau_2}|(2l)+(2k+1)\tau|^2\right]
    Z_L^{(1,0)}(\tau){Z^{2k}_{1+2l}(\tau)^*}^4 && \nonumber \\
    \left.
   +\sum_{k,l\in{\bf Z}}\exp\left[-\frac{\pi R^2}{\alpha'\tau_2}|(2l+1)+(2k+1)\tau|^2\right]
    Z_L^{(1,1)}(\tau){Z^{2k}_{2l}(\tau)^*}^4 \right\}. && \nonumber \\
\end{eqnarray}
We have replaced $|Z^{1+2mqR}_{1+2nqR}(\tau)|^{-2}$ in the eq.(\ref{fullPF}), which is 
singular at $q=1/R$, with $\tau_2^{-1}|\eta(\tau)|^{-4}$. 
From the Poisson resummation formula \cite{CostaGutperle}, 
\begin{equation}
\sum_{k,l\in{\bf Z}}\exp\left[-\frac{\pi R^2}{\alpha'\tau_2}
                              |(2l+\delta)+(2k+\epsilon)\tau|^2\right]
= {\cal Z}_{(\epsilon,0)}+(-1)^\delta{\cal Z}_{(\epsilon,1)}, 
\end{equation}
where
\begin{equation}
{\cal Z}_{(\epsilon,\delta)} = \frac{\sqrt{\alpha'\tau_2}}{2R}
\sum_{k,l\in{\bf Z}}\exp\left[-\pi\tau_2\left\{\frac{\alpha'}{4R^2}(2l+\delta)^2
  +\frac{4R^2}{\alpha'}(k+\epsilon/2)^2\right\}+2\pi i\tau_1(2l+\delta)(k+\epsilon/2)\right]. 
\end{equation}
Then we find that
\begin{eqnarray}
Z_{1/R}(A_Y^I) 
= \int \frac{d^2\tau}{\tau_2}\tau_2^{-5}|\eta(\tau)|^{-16} \left\{
 {\cal Z}_{(0,0)}\left(Z_L^{(0,0)}(\tau){Z^1_1(\tau)^*}^4+Z_L^{(0,1)}(\tau){Z^1_0(\tau)^*}^4
 \right)\right. && \nonumber \\
+{\cal Z}_{(0,1)}\left(Z_L^{(0,0)}(\tau){Z^1_1(\tau)^*}^4-Z_L^{(0,1)}(\tau){Z^1_0(\tau)^*}^4
 \right) && \nonumber \\
+{\cal Z}_{(1,0)}\left(Z_L^{(1,0)}(\tau){Z^0_1(\tau)^*}^4+Z_L^{(1,1)}(\tau){Z^0_0(\tau)^*}^4
 \right) && \nonumber \\
 \left.
+{\cal Z}_{(1,1)}\left(Z_L^{(1,0)}(\tau){Z^0_1(\tau)^*}^4-Z_L^{(1,1)}(\tau){Z^0_0(\tau)^*}^4
 \right) \right\} &&.  \nonumber \\ 
\end{eqnarray}
Surprisingly, this partition function can be interpreted as that of an orbifold. 
The orbifold can be constructed from the supersymmetric heterotic theory by the ${\bf Z}_2$ 
twist whose generator is $(-1)^{F_s}\gamma_\delta\ \sigma_{1/2}$. 
Here $\sigma_{1/2}$ is the half-shift operator \cite{Type0}, and the other two are the 
same operators as the ones explained in section 4. 
The sum (\ref{sum}) is equal to the length squared of the shift vector $\delta$, and the 
requirement of the 
periodicity (\ref{periodicity}) is equivalent to the idenpotency of $\gamma_\delta$, i.e. 
$\gamma_\delta^2=1$. 

\vspace{2mm}

Consider the $R\to 0$ limit. 
Then ${\cal Z}_{(0,1)}$ and ${\cal Z}_{(1,1)}$ vanish, while the other two sums approach 
the same non-zero value which is independent of $\tau_2$. 
What survive this limit are the followings,
\begin{eqnarray}
Z_{1/R}(A_Y^I) 
&\to& \int \frac{d^2\tau}{\tau_2}\tau_2^{-5}|\eta(\tau)|^{-16} \left\{
 \left(Z_L^{(0,0)}(\tau){Z^1_1(\tau)^*}^4+Z_L^{(0,1)}(\tau){Z^1_0(\tau)^*}^4
 \right) \right. \nonumber \\
& & \hspace{3cm}\left.
+\left(Z_L^{(1,0)}(\tau){Z^0_1(\tau)^*}^4+Z_L^{(1,1)}(\tau){Z^0_0(\tau)^*}^4
 \right) \right\} .
\end{eqnarray}
This is the partition function for a non-supersymmetric heterotic theory compactified on 
the vanishing $S^1$. 

Thus we conclude that a supersymmetric heterotic theory on the critical Melvin background 
with the appropriate Wilson line has the same mass spectrum as that of a non-supersymmetric 
heterotic theory, in the $R\to 0$ limit. 
This result can be understood, in the T-dual picture, as evidence of the equivalence between 
the supersymmetric theory on a non-trivial background and a non-supersymmetric theory
in ten dimensions, as in the IIA-0A case. 

\vspace{2mm}

The result for the case with odd values of the sum (\ref{sum}) is almost the same. 
The only difference is the sign in front of ${Z^0_0(\tau)^*}^4$. 
The change of this sign corresponds to the fact that the eigenvalues of 
$(-1)^{F_s}\gamma_\delta$ are a bit peculiar in the twisted sectors, as mentioned in 
\cite{nonSUSYhet}. 

\vspace{2mm}

We have constructed all non-supersymmetric heterotic theories but one in terms of the 
supersymmetric heterotic theories. 
The explicit expressions of the partition functions are shown in the appendix. 
The only exception is the theory whose gauge group is $E_8$. 
It is argued in \cite{nonSUSYhet} that to obtain the heterotic $E_8$ theory one has to 
twist 
the $E_8\times E_8$ theory by a ${\bf Z}_2$ involving the outer automorphism of 
$E_8\times E_8$ which exchanges the two $E_8$'s. 
We will discuss this theory in section 8.

\vspace{1cm}

\section{Tachyonic instability}

\vspace{5mm}

In the IIA-0A case, the tachyonic instability is interpreted as the one due to pair 
creations of D6-branes. 
This argument is based on the analysis of the instability of the Melvin background. 
We will discuss in this section the instabiliity of the non-supersymmetric heterotic 
theories in terms of the background fields in the supersymmetric theories.

\vspace{5mm}

\subsection{Heterotic T-duality}

\vspace{3mm}

We have considered the heterotic theories on the Kaluza-Klein Melvin background with a 
vanishing 
radius of the $S^1$. 
So it is not appropriate for the argument on the background fields. 
The ordinary way to improve this situation is to see the T-dual picture. 

Let us recall the T-duality transformation for the heterotic theories. 
This can be most easily derived from the sigma-model approach \cite{T-duality}. 

The sigma-model action of the heterotic theories in a general background is written by 
superfields as follows,
\begin{eqnarray}
S &=& \frac i{2\pi \alpha'}\int d^2\sigma d\theta
     \left\{G_{mn}(\Phi)+B_{mn}(\Phi)\right\}\partial_+\Phi^m D_-\Phi^n
   \nonumber \\
&&  +\frac1{2\pi}\int d^2\sigma d\theta
     \Psi^I\left\{D_-\Psi^I-iA_m^{IJ}(\Phi)D_-\Phi^m\Psi^J\right\},
  \label{SigmaAction}
\end{eqnarray}
where $\theta$ is the Grassmann coordinate, $m,n=0,\cdots,9$ and $I,J=1,\cdots,32$. 
We have defined the real bosonic superfields $\Phi^m$ and real fermionic ones $\Psi^I$ as 
follows,
\begin{eqnarray}
&& \Phi^m = X^m+i\sqrt{2\alpha'}\theta\psi_R^m, \\
&& \Psi^I = \lambda_L^I+\theta F^I,
\end{eqnarray}
where we have employed the RNS formalism for both left- and right-moving fermions. 
The covariant derivative $D_-$ is 
\begin{equation}
D_- = \partial_\theta -2i\theta\partial_-\ .
\end{equation}
After the integration of $\theta$ and integrating out the auxiliary fields $F^I$, the action 
(\ref{SigmaAction}) becomes
\begin{eqnarray}
S &=& \int d^2\sigma\left[
 \frac1{\pi\alpha'}\left\{G_{mn}(X)+B_{mn}(X)\right\}\partial_+X^m\partial_-X^n
  \right.  \nonumber \\
& & \hspace*{1cm} +\frac i{\pi}G_{mn}(X)\psi_R^m
  \left\{\partial_+\psi_R^n+\frac12(\Gamma^n_{kl}(X)+H^n_{kl}(X))
  \partial_+X^k\psi_R^l\right\} \nonumber \\
& & \left. \hspace*{1cm}
   +\frac i{\pi}\lambda_L^I\left\{\partial_-\lambda^I-iA_m^{IJ}(X)\partial_-X^m
     \lambda_L^J\right\}
   -i\alpha'F_{mn}^{IJ}(X)\lambda_L^I\lambda_L^J\psi_R^m\psi_R^n\right]\ .
\end{eqnarray}
The action (\ref{HetAction}) is obtained by substituting the metric of the Melvin background 
and the 
constant gauge fields. 

Now suppose that the background fields are independent of a coordinate, say, $x^9$, or in 
other words there is an isometry along the 9-direction. 
Then there exists a dual sigma-model which is equivalent to the original model. 
This can be shown by considering the following first order action, 
\begin{eqnarray}
S_{1st} &=& \int d^2\sigma d\theta\left[
  \frac i{2\pi\alpha'}\left( \{G_{ij}+B_{ij}\}\partial_+\Phi^iD_-\Phi^j
                            +i\{G_{i9}+B_{i9}\}\partial_+\Phi^iV_-
   \right. \right. \nonumber \\
&& \left.  \hspace*{1cm}    +\{G_{9j}+B_{9j}\}V_+D_-\Phi^j
                            +iG_{99}V_+V_- \right)
   \nonumber \\
&& \hspace*{1cm}
   +\frac1{2\pi}\left( \Psi^I(D_-\Psi^I-iA_i^{IJ}D_-\Phi^i\Psi^J)
                     +\Psi^IA_9^{IJ}V_-\Psi^J \right)
   \nonumber \\
&& \hspace*{1cm}
   \left. +\tilde{\Phi}(i\partial_+V_--D_-V_+) \right]
  \label{1stAction}
\end{eqnarray}
where $i,j=0,\cdots,8$. 
We have introduced a bosonic and a fermionic superfield $V_+$, $V_-$ respectively, and 
$\tilde{\Phi}$ is a bosonic Lagrange multiplier superfield. 

When the $\tilde{\Phi}$ integration is performed, the superfields $V_+,\ V_-$ are determined 
to be 
\begin{equation}
V_+=\partial_+\Phi^9,\hspace{1cm} V_-=-iD_-\Phi^9,
\end{equation}
for a real bosonic superfield $\Phi^9$. 
Then the action (\ref{1stAction}) becomes the original action (\ref{SigmaAction}). 

On the other hand, when one integrates out $V_+,\ V_-$ first, which can be done exactly, 
one obtains a sigma-model with the background fields 
$\tilde{G}_{\mu\nu},\ \tilde{B}_{\mu\nu},\ \tilde{A}_\mu^{IJ}$ such that \cite{T-duality}
\begin{eqnarray}
&& \tilde{G}_{ij} = G_{ij}-\frac1{G_{99}}(G_{i9}G_{9j}+B_{i9}B_{9j}), \hspace{5mm}
   \tilde{B}_{ij} = B_{ij}-\frac1{G_{99}}(G_{i9}B_{9j}-G_{j9}B_{9i}), \nonumber \\
&& \tilde{G}_{i9} = -\frac1{G_{99}}B_{i9}, \hspace{5mm}
   \tilde{B}_{i9} = -\frac1{G_{99}}G_{i9}, \hspace{5mm}
   \tilde{G}_{99} = \frac1{G_{99}}, 
  \label{Ttransf} \\
&& \tilde{A}_i^{IJ} = A_i^{IJ}-\frac1{G_{99}}(G_{9i}+B_{9i})A_9^{IJ},\hspace{5mm}
   \tilde{A}_9^{IJ} = -\frac1{G_{99}}A_9^{IJ}\ . \nonumber 
\end{eqnarray}
We have set $\alpha'=1$. 
Note that the dilaton field $\phi$ is also transformed as 
\begin{equation}
\tilde{\phi} = \phi-\frac12\log G_{99}. 
\end{equation}

\vspace{5mm}

\subsection{T-dual of Melvin background}

\vspace{3mm}

We can now consider the T-dual of the Melvin background. 
First let us rescale the $y$-coordinate so that its period is $2\pi$. 
Then applying the transformation rules (\ref{Ttransf}), one obtains 
\begin{eqnarray}
ds_{dual}^2 &=& \eta_{\mu\nu}dx^\mu dx^\nu+dr^2+\frac{r^2}{1+q^2r^2}d\theta^2
               +\frac1{R^2(1+q^2r^2)}dy^2, \\
\tilde{B}_{\theta y} &=& -\frac{qr^2}{R(1+q^2r^2)}, \\
\tilde{A}_\theta^{IJ} &=& -\frac{qr^2}{R^2(1+q^2r^2)}a^{IJ}, \hspace{5mm}
\tilde{A}_y^{IJ} \ =\  -\frac1{R^3(1+q^2r^2)}a^{IJ},
\end{eqnarray}
where $a^{IJ}=RA_y^{IJ}$ is independent of both $R$ and $q$. 
The region $r\ll q^{-1}$ is almost flat and the radius of the $S^1$ is large, so this 
background 
would be reasonable in this region. 
An important property of this background is that there exists a nontrivial Yang-Mills field 
strength, 
\begin{eqnarray}
&& \tilde{F}_{78}^{IJ} = -\frac{2qa^{IJ}}{R^2(1+q^2r^2)^2}, \\
&& \tilde{F}_{89}^{IJ} = \ \ \frac{2q^2a^{IJ}}{R^3(1+q^2r^2)^2}x^8, \\
&& \tilde{F}_{97}^{IJ} = -\frac{2q^2a^{IJ}}{R^3(1+q^2r^2)^2}x^7.
\end{eqnarray}
In particular, around the center $r=0$, there is a magnetic field $\tilde{F}^{IJ}_{78}$. 
This magnetic field would cause, for example, monopole pair creations, and this system would 
decay into a stable background. 
This might be the dual picture of the closed string tachyon condensation in the 
non-supersymmetric heterotic theories. 
It seems natural to relate the tachyonic instability to the existence of the Yang-Mills 
magnetic 
field since the closed string tachyons are, in general, in a nontrivial representation of 
the gauge group. 

Note that for large $r$, $G_{\theta\theta}$ approaches $q^{-2}$. 
So the analysis based on the background wolud not be valid when $q$ is large. 
Thus the discussion given above could not apply to the critical Melvin background, i.e. 
the case 
with $q=1/R$. 
Nevertheless, we expect that the above-mentioned property still holds at the critical value 
of 
$q$. 

\vspace{2mm}

It is interesting to consider the case in which there is no Wilson line in the original 
background. 
Then the gauge field is again absent in the T-dual picture. 
So there is no instability due to the Yang-Mills field strength. 
In fact, the situation is very close to the IIA-0A case. 
It is discussed in \cite{bubble} that the background is stable if there exists a fermion. 
In Type 0A theory, there is no spacetime fermion and the background is unstable due to the 
bubble nucleation \cite{bubble}. 
On the other hand, in the heterotic theories there are many spacetime fermions, so this may 
indicate that the critical Melvin background is stable in this case. 
This is consistent with the fact that the partition function vanishes at $q=1/R$ if there is 
no 
Wilson line, suggesting that the theory becomes supersymmetric.

\vspace{1cm}

\section{Strong coupling behavior}

\vspace{5mm}

In this section, we will investigate the strong coupling region of the non-supersymmetric 
heterotic theories. 
The discussions so far are mainly based on the weakly-coupled string theory, and we have not 
discussed quantum corrections at all. 
Because of the absence of the supersymmetry, there is no argument which ensures that the 
tree-level analysis is sufficient.  
There are several arguments for the dual theory of non-supersymmetric theories 
\cite{Type0}\cite{0dual}\cite{Blum}. 

Our starting point is the existence of the dual picture shown in section 5. 
The advantage to use this fact is that the duality transformations of the supersymmetric 
theory 
are available to investigate the dual of a non-supersymmetric theory. 
So our arguments would be based on a rather firmer footing, even if there is no 
supersymmetry. 

Consider a non-supersymmetric heterotic theory compactified on $S^1$ with radius $R$. 
Denote its coupling constant as $g$. 
Its strong coupling behavior would be described by a strongly-coupled supersymmetric 
heterotic 
theory on the critical Melvin background, in the $R\to 0$ limit. 
The latter theory is either the $SO(32)$ theory or the $E_8\times E_8$ theory, according to 
which theory we consider as the former.

\vspace{5mm}

(i) The $SO(32)$ theory

\vspace{5mm}

What we consider now is the strongly-coupled $SO(32)$ theory on a nontrivial background. 
Its weak coupling dual is well-known \cite{PolWitten} to be Type I theory on the following 
background. 
\begin{eqnarray}
&& G_{I,mn} = e^{-\phi_h}G_{h,mn} \\
&& \phi_I = -\phi_h \\
&& A_{I,m} = A_{h,m}
\end{eqnarray}
The subscripts $I,h$ indicate Type I theory and the heterotic theory, respectively. 
Note that in our case, there is no R-R background in the dual Type I background. 
The dilaton $\phi_h$ is a constant, so that the Type I metric is also the Melvin background. 

Type I theory can be understood as an orientifold of Type IIB theory \cite{Sagnotti}. 
So our dual theory is the orientifold of Type IIB theory on the critical Melvin. 
It can be shown that Type IIB theory on the critical Melvin has the same partition function 
as 
that of Type 0B theory in the $R\to0$ limit, just as in the IIA-0A case. 
Therefore the dual theory can be further reinterpreted as an orientifold of Type 0B theory. 

\vspace{2mm}

Interestingly enough, this orientifold is what was proposed as the dual theory of the 
$SO(16)\times SO(16)$ heterotic theory \cite{Blum}. 
More precisely, in \cite{Blum} the dual theory is constructed as an orientifold of an 
interpolating model which relates Type IIB theory to Type 0B theory. 
The partition function of the interpolating model coincides with that of Type IIB theory on 
the critical Melvin with a finite radius. 
So the dual theory proposed in \cite{Blum} is the same as our dual theory. 
Note that the discussion in \cite{Blum} is based on deformations of the supersymmetric 
theories, and it is not obvious whether the Type I-heterotic duality can be applied in this 
case, as mentioned by the authers. 
Since we have related the non-supersymmetric theories to the supersymmetric theories on 
some background, there would be no obstruction to apply the duality relation. 

The pieces of evidence for the Type 0B orientifold to be the dual of the non- supersymmetric 
$SO(16)\times SO(16)$ theory are the followings \cite{Blum}: 

\vspace{1mm}

 1. The massless spectrum is the same. 

 2. It is argued that fluctuations on the D1-brane in the Type 0B orientifold corresponds 
\hspace*{1cm}to the worldsheet fields in non-supersymmetric theory, under a few assumptions. 

\vspace{1mm}

These are nice properties for them to be a dual pair. 
It will be interesting to confirm the discussions in \cite{Blum} from our point of view based 
on the supersymmetric theory. 

\vspace{2mm}

Since in the Type 0B orientifold picture the radius of the $y$-direction is still 
small, we have to take the T-dual transformation along the $y$-direction. 
Then we obtain Type 0A theory compactified on $S^1/{\bf Z}_2$. 
It would be natural to expect that the Type 0A orbifold decays into a Type IIA orbifold, just 
like the decay in the IIA-0A case in ten dimensions. 
Note that the decay process would be different from the IIA-0A case, in particular, in the 
Type 
0A orbifold which is related to the $SO(16)\times SO(16)$ heterotic theory, since the 
orbifold does not contain any tachyon. 
It is well-known that the Type IIA orbifold is the dual of a supersymmetric heterotic theory 
(with an appropriate Wilson line). 
Therefore we conclude that the picture of the decay of the non-supersymmetric theories 
obtained 
from the perturbative analysis still hold in the strong coupling region. 

\vspace{2mm}

Let us determine the coupling constant $g_{0A}$ and the length $R_{0A}$ of the segment 
$S^1/{\bf Z}_2$ in the Type 0A orbifold when $g$ is large and $R$ is small in the 
corresponding non-supersymmertic theory. 
Applying the succesive duality transformations, one finds 
\begin{equation}
g_{0A} \sim g^{-\frac12}R^{-1}, \hspace{1cm} R_{0A} \sim g^{\frac12}R^{-1},
\end{equation}
up to numerical factors. 

For large but finite $g$, the segmant is large but the coupling constant also becomes large 
when $R$ is small enough. 
This would not be so problematic since the arguments on the IIA-0A case are not restricted in 
the perturbative region. 
One can obtain, if necessary, a weakly-coupled Type 0A orbifold as the dual theory, by taking 
the 
following limit. 
\begin{equation}
g\to \infty, \hspace{7mm} R\to 0, \hspace{7mm} gR^2:\mbox{large but finite}
\end{equation}

\vspace{5mm}

(ii) The $E_8\times E_8$ theory

\vspace{3mm}

The dual of the strongly-coupled $E_8\times E_8$ theory on the critical Melvin is rather easy 
to obtain. 
It is M-theory on ${\bf R}^9\times S^1\times S^1/{\bf Z}_2$, with the anti-periodic boundary 
condition for fermions along the $S^1$. 
After reducing M-theory along the $S^1/{\bf Z}_2$, 
this corresponds to the non-supersymmetric heterotic theory in the $R\to 0$ limit, where 
$R$ is the radius of the $S^1$. 
Then, exchanging the roles of the $S^1$ and the $S^1/{\bf Z}_2$, and reducing the theory 
along 
the $S^1$ direction, we obtain a weakly-coupled Type 0A theory on the $S^1/{\bf Z}_2$. 
This is because M-theory compactified on such a twisted circle corresponds to Type 0A 
theory \cite{Type0}, which is one of the main point of the IIA-0A duality. 
Thus we again find the Type 0A orbifold as the dual theory of the non-supersymmetric 
heterotic 
theories. 
So the non-supersymmetric theories, which are related to the $E_8\times E_8$ theory, are 
expected to decay into the supersymmetric theory even in the strong coupling region.

\vspace{1cm}

\section{Discussion}

\vspace{5mm}

We have constructed the non-supersymmetric heterotic theories as the supersymmetric 
heterotic theories on the critical Melvin background with the Wilson lines. 
However, we have not discussed such a construction of the $E_8$ theory. 
This theory is rather exceptional; for example, the rank of its gauge group is eight, while 
all the other theories have the gauge group of the rank sixteen. 
Moreover, to construct the $E_8$ theory via the twisting procedure from the $E_8\times E_8$ 
theory, the ${\bf Z}_2$ group should consist of the outer automorphism of $E_8\times E_8$ 
which exchanges the two $E_8$'s. 
So this theory does not seem to be constructed as a theory on the critical Melvin. 
One possible supersymmetric theory which can be related to this theory is the CHL theory 
\cite{CHL}. 
The CHL theory can have the gauge group of the rank eight, and it can be constructed by 
twisting the $E_8\times E_8$ theory by a ${\bf Z}_2$ which consists of the outer 
automorphism discussed above. 
Since the CHL theory is supersymmetric, it would be natural to expect that the $E_8$ theory 
decays into this theory. 

The non-supersymmetric theories have been related to the critical Melvin 
background in the supersymmetric theories. 
This background has nontrivial field strength, both in the lower-dimensional and T-dual 
point of view. 
Such configurations have nonzero energy, compared with the supersymmetric vacuum. 
So one is led to a conjecture which suggests that the ``vacuum energy'' of the 
non-supersymmetric 
theories are, in general, positive and their values are given by the energy of the field 
strength in the dual supersymmetric theories. 
This might have something to do with the fact that the one loop cosmological constant of the 
$SO(16)\times SO(16)$ theory is positive \cite{nonSUSYhet}. 

In section 7, we have discussed the strong coupling behavior of the non-supersymmetric 
theories. 
This kind of analysis is in general very difficult because of the absence of supersymmetry. 
In our case, it can be done since the non-supersymmetric theories are related to the 
non-supersymmetric backgrounds in the supersymmetric theories, and the duality 
transformations for the background fields are well-known. 
This seems to be a very suitable strategy to discuss the duality between non-supersymmetric 
theories. 
It would be very interesting if this strategy can apply to some other non-supersymmetric 
theories. 

One of our motivations to study the Melvin background in string theory is to understand 
the mechanism of the stabilization of closed string tachyons. 
In \cite{CostaGutperle}\cite{fluxbrane}, the tachyon condensation is related to pair 
creations of D6-branes in Type IIA picture. 
However, the corresponding picture in Type 0A theory is not yet clarified. 
The bubble nucleation in \cite{bubble} pushes our world to infinity, and there remains 
nothing. 
It would be very important to understand this phenomenon in terms of Type 0A theory 
language, and clarify the mechanism of the closed string tachyon condensation. 

In the heterotic theories, the situation is more complicated. 
Since most of the theories have tachyons in a nontrivial representation of their gauge 
groups, the condensation of the tachyons would break the gauge symmetry. 
Therefore, even if the endpoint of the condensation is a supersymmetric theory, 
there should be some other changes of the background, for example, the change of the pattern 
of the Wilson line. 
This might suggest that the tachyon condensation in heterotic theories could not be 
described by the change in $q$ alone.  
 
\vspace{1cm}

{\Large {\bf Acknowledgements}}

\vspace{5mm}

I would like to thank S. Iso, H. Itoyama and T. Matsuo for valuable discussions. 

\newpage

\appendix

\vspace{1cm}

\section{Explicit expressions of partition functions}

\vspace{5mm}

In this appendix, we will show explicitly the partition functions $Z_{1/R}(A_y^I)$ discussed 
in section 5. 
First we have to choose the appropriate Wilson lines \cite{nonSUSYhet}. 
We assumed the periodicity of the left-moving contribution (\ref{periodicity}). 
This means that the sixteen dimensional vector 
\begin{equation}
\delta = (a^1,\cdots,a^{16})
\end{equation}
has the property that $2\delta$ lies in the momentum lattice, for a suitable choice of the 
basis vectors. 
Such vectors are classified in \cite{WilsonLine} and the followings are the 
representatives of the equivalence classes. 
For the $SO(32)$ theory, 
\begin{equation}
\delta = (1,\ 0^{15}),\ \left(\left(\mbox{$\frac12$}\right)^4,\ 0^{12}\right), \ 
         \left(\left(\mbox{$\frac14$}\right)^{16}\right), \ 
           \left(\left(\mbox{$\frac12$}\right)^8,\ 0^8\right),
\end{equation}
and for the $E_8\times E_8$ theory, 
\begin{equation}
\delta = (1,\ 0^7;\ 0^8),\ 
           \left(\left(\mbox{$\frac12$}\right)^2, \ 0^6;\ 
                 \left(\mbox{$\frac12$}\right)^2, \ 0^6\right), \ 
        (1,\ 0^7;\ 1,\ 0^7).
\end{equation}

Below we denote the partition functions for various sectors as 
\begin{eqnarray}
\chi_{NS\pm}^{(n)} &=& \frac12\left(Z^0_0(\tau)^n\pm Z^0_1(\tau)^n\right), \\
\chi_{R}^{(n)} &=& \frac12Z^1_0(\tau)^n,
\end{eqnarray}
for the left-movers and
\begin{eqnarray}
\bar{\chi}_{NS\pm} &=& \frac12\left(Z^0_0(\tau)^4\mp Z^0_1(\tau)^4\right)^*, \\
\bar{\chi}_{R} &=& \frac12{Z^1_0(\tau)^4}^*
\end{eqnarray}
for the right-movers.

\vspace{5mm}

\subsection{The $SO(32)$ theory}

\vspace{3mm}

(i) $\delta = (1,\ 0^{15})$

\vspace{2mm}

In this case, the partition function $Z_L^{(m,n)}(\tau)$ for the left-movers is 
\begin{equation}
Z_L^{(m,n)}(\tau) = Z^0_0(\tau)^{16}+e^{-\pi im}Z^0_1(\tau)^{16}+e^{\pi in}Z^1_0(\tau)^{16}.
\end{equation}
Then the full partition function is 
\begin{equation}
Z_{1/R}(A_y^I) = \int \frac{d^2\tau}{\tau_2}\tau_2^{-5}|\eta(\tau)|^{-16}Z_F,
\end{equation}
where 
\begin{eqnarray}
Z_F &=& 4\left\{\ \ 
  {\cal Z}_{(0,0)}(\chi_{NS+}^{(16)}\bar{\chi}_{NS+}
                  -\chi_{R}^{(16)}\bar{\chi}_{R})
 +{\cal Z}_{(0,1)}(\chi_{R}^{(16)}\bar{\chi}_{NS+}
                  -\chi_{NS+}^{(16)}\bar{\chi}_{R}) \right. \nonumber \\
    & & \hspace*{5mm} \left.
 +{\cal Z}_{(1,0)}(\chi_{NS-}^{(16)}\bar{\chi}_{NS-}
                  -\chi_{R}^{(16)}\bar{\chi}_{R})
 +{\cal Z}_{(0,1)}(\chi_{R}^{(16)}\bar{\chi}_{NS-}
                  -\chi_{NS-}^{(16)}\bar{\chi}_{R}) \right\}.
\end{eqnarray}
In the $R\to 0$ limit, 
\begin{equation}
Z_F \ \propto \ \chi_{NS+}^{(16)}\bar{\chi}_{NS+}
           +\chi_{NS-}^{(16)}\bar{\chi}_{NS-}
           -2\chi_{R}^{(16)}\bar{\chi}_{R}.
\end{equation}
So $Z_{1/R}(A_y^I)$ coincides exactly with the partition function for the non-supersymmetric 
$SO(32)$ theory in this limit. 

\vspace{6mm}

(ii) $\delta = \left(\left(\mbox{$\frac12$}\right)^4,\ 0^{12}\right)$

\vspace{2mm}

In this case, from the expression
\begin{equation}
Z_L^{(m,n)}(\tau) = Z^m_n(\tau)^4Z^0_0(\tau)^{12}+Z^m_{n+1}(\tau)^4Z^0_1(\tau)^{12}
                   +Z^{m+1}_n(\tau)^4Z^1_0(\tau)^{12},
\end{equation}
one can find 
\begin{eqnarray}
Z_F &=& 4\left[\ \ 
  {\cal Z}_{(0,0)}\left\{(\chi_{NS+}^{(4)}\chi_{NS+}^{(12)}
                         +\chi_{R}^{(4)}\chi_{R}^{(12)})
                          \bar{\chi}_{NS+}
                        -(\chi_{NS-}^{(4)}\chi_{NS-}^{(12)}
                         +\chi_{R}^{(4)}\chi_{R}^{(12)})
                          \bar{\chi}_{R}\right\} \right. \nonumber \\
    & & \hspace*{5mm}
 +{\cal Z}_{(0,1)}\left\{(\chi_{NS-}^{(4)}\chi_{NS-}^{(12)}
                         +\chi_{R}^{(4)}\chi_{R}^{(12)})
                          \bar{\chi}_{NS+}
                        -(\chi_{NS+}^{(4)}\chi_{NS+}^{(12)}
                         +\chi_{R}^{(4)}\chi_{R}^{(12)})
                          \bar{\chi}_{R}\right\}  \nonumber \\
    & & \hspace*{5mm}
 +{\cal Z}_{(1,0)}\left\{(\chi_{R}^{(4)}\chi_{NS+}^{(12)}
                         +\chi_{NS+}^{(4)}\chi_{R}^{(12)})
                          \bar{\chi}_{NS-}
                        -(\chi_{R}^{(4)}\chi_{NS-}^{(12)}
                         +\chi_{NS-}^{(4)}\chi_{R}^{(12)})
                          \bar{\chi}_{R}\right\} \nonumber \\
    & & \hspace*{5mm} \left.
 +{\cal Z}_{(1,1)}\left\{(\chi_{R}^{(4)}\chi_{NS-}^{(12)}
                         +\chi_{NS-}^{(4)}\chi_{R}^{(12)})
                          \bar{\chi}_{NS-}
                        -(\chi_{R}^{(4)}\chi_{NS+}^{(12)}
                         +\chi_{NS+}^{(4)}\chi_{R}^{(12)})
                          \bar{\chi}_{R}\right\} \right] . \nonumber \\
\end{eqnarray}
In the $R\to 0$ limit, 
\begin{eqnarray}
Z_F &\propto&   (\chi_{NS+}^{(4)}\chi_{NS+}^{(12)}
                         +\chi_{R}^{(4)}\chi_{R}^{(12)})
                          \bar{\chi}_{NS+}
                        -(\chi_{NS-}^{(4)}\chi_{NS-}^{(12)}
                         +\chi_{R}^{(4)}\chi_{R}^{(12)})
                          \bar{\chi}_{R}
                \nonumber \\
   & &         +(\chi_{R}^{(4)}\chi_{NS+}^{(12)}
                         +\chi_{NS+}^{(4)}\chi_{R}^{(12)})
                          \bar{\chi}_{NS-}
                        -(\chi_{R}^{(4)}\chi_{NS-}^{(12)}
                         +\chi_{NS-}^{(4)}\chi_{R}^{(12)})
                          \bar{\chi}_{R}.
\end{eqnarray}

The corresponding shift operator $\gamma_\delta$ can be represented as 
\begin{eqnarray}
\gamma_\delta &=& e^{\pi i(J_{12}+J_{34}+J_{56}+J_{78})} \nonumber \\
              &=& (-1)^{F^{(4)}},
\end{eqnarray}
where $F^{(4)}$ counts the number of $\lambda_L^1,\cdots,\lambda_L^4$. 
Therefore one can see that the partition function in the $R\to 0$ limit coincides with that 
of the supersymmetric $SO(32)$ theory twisted by $(-1)^{F_s}\gamma_\delta$, which 
is the $SO(8)\times SO(24)$ theory. 

\vspace{6mm}

(iii) $\delta = \left(\left(\mbox{$\frac14$}\right)^{16}\right)$

\vspace{2mm}

The expression for the left-moving contribution is 
\begin{equation}
Z_L^{(m,n)}(\tau) = Z^{m/2}_{n/2}(\tau)^{16}+Z^{m/2}_{1+n/2}(\tau)^{16}
                   +Z^{1+m/2}_{n/2}(\tau)^{16}+Z^{1+m/2}_{1+n/2}(\tau)^{16}. 
\end{equation}
The periodicity (\ref{periodicity}) is accomplished by permuting the terms in the RHS. 

The partition function is given by 
\begin{eqnarray}
Z_F &=& 4\left[\ \ 
  {\cal Z}_{(0,0)}\left\{(\chi_{NS+}^++\chi_{R+}^+)\bar{\chi}_{NS+}
                        -(\chi_{NS+}^-+\chi_{R+}^-)\bar{\chi}_R\right\} \right.
    \nonumber \\
    & & \hspace*{5mm}
 +{\cal Z}_{(0,1)}\left\{(\chi_{NS+}^-+\chi_{R+}^-)\bar{\chi}_{NS+}
                        -(\chi_{NS+}^++\chi_{R+}^+)\bar{\chi}_R\right\} \nonumber \\
    & & \hspace*{5mm}
 +{\cal Z}_{(1,0)}\left\{(\tilde{\chi}_{NS+}^++\tilde{\chi}_{R+}^+)\bar{\chi}_{NS-}
                        -(\tilde{\chi}_{NS+}^-+\tilde{\chi}_{R+}^-)\bar{\chi}_R\right\} 
    \nonumber \\
    & & \hspace*{5mm} \left.
 +{\cal Z}_{(1,1)}\left\{(\tilde{\chi}_{NS+}^-+\tilde{\chi}_{R+}^-)\bar{\chi}_{NS-}
                        -(\tilde{\chi}_{NS+}^++\tilde{\chi}_{R+}^+)\bar{\chi}_R\right\} 
    \right],
\end{eqnarray}
where 
\begin{eqnarray}
   \chi_{NS+}^\pm &=& \frac14\left(
        Z^0_0(\tau)^{16}+Z^0_1(\tau)^{16}
    \pm(Z^0_{\frac12}(\tau)^{16}+Z^0_{1+\frac12}(\tau)^{16})\right),
   \\
   \chi_{R+}^\pm &=& \frac14\left(
        Z^1_0(\tau)^{16}+Z^1_1(\tau)^{16}
    \pm(Z^1_{\frac12}(\tau)^{16}+Z^1_{1+\frac12}(\tau)^{16})\right),
   \\
   \tilde{\chi}_{NS+}^\pm &=& \frac14\left(
        Z^{\frac12}_0(\tau)^{16}+Z^{\frac12}_1(\tau)^{16}
    \pm(Z^{\frac12}_{\frac12}(\tau)^{16}+Z^{\frac12}_{1+\frac12}(\tau)^{16})\right),
   \\
   \tilde{\chi}_{R+}^\pm &=& \frac14\left(
        Z^{1+\frac12}_0(\tau)^{16}+Z^{1+\frac12}_1(\tau)^{16}
    \pm(Z^{1+\frac12}_{\frac12}(\tau)^{16}+Z^{1+\frac12}_{1+\frac12}(\tau)^{16})\right).
\end{eqnarray}
Note that $\chi_{NS+(R+)}^\pm$ are the projected partition functions
\begin{equation}
\chi_{NS+(R+)}^\pm = \mbox{Tr}_{NS(R)}\frac{1+(-1)^F}2\frac{1\pm \gamma_\delta}2q^H, 
\end{equation}
where $\gamma_\delta$ acts as $e^{\frac{\pi}2i}$ on all $\lambda_L^I$'s. 
The other two functions $\tilde{\chi}_{NS+(R+)}^\pm$ are the projected partition functions 
for the twisted sectors. 

In the $R\to 0$ limit, 
\begin{eqnarray}
Z_F &\propto& \ \ 
              (\chi_{NS+}^++\chi_{R+}^+)\bar{\chi}_{NS+}
             -(\chi_{NS+}^-+\chi_{R+}^-)\bar{\chi}_{R} \nonumber \\
    & &      +(\tilde{\chi}_{NS+}^++\tilde{\chi}_{R+}^+)\bar{\chi}_{NS-}
             -(\tilde{\chi}_{NS+}^-+\tilde{\chi}_{R+}^-)\bar{\chi}_{R}. 
\end{eqnarray}
This is the fermion part of the partition function for the non-supersymmetric $U(16)$ theory. 

\vspace{6mm}

(iv) $\delta = \left(\left(\mbox{$\frac12$}\right)^8,\ 0^8\right)$

\vspace{2mm}

The calculation is similar to the one in the case (ii) above, except for the extra signs in 
the twisted sectors. 
The partition function is then given by

\begin{eqnarray}
Z_F &=& 4\left[\ \ 
  {\cal Z}_{(0,0)}\left\{(\chi_{NS+}^{(8)}\chi_{NS+}^{(8)}
                         +\chi_{R}^{(8)}\chi_{R}^{(8)})
                          \bar{\chi}_{NS+}
                        -(\chi_{NS-}^{(8)}\chi_{NS-}^{(8)}
                         +\chi_{R}^{(8)}\chi_{R}^{(8)})
                          \bar{\chi}_{R}\right\} \right. \nonumber \\
    & & \hspace*{5mm}
 +{\cal Z}_{(0,1)}\left\{(\chi_{NS-}^{(8)}\chi_{NS-}^{(8)}
                         +\chi_{R}^{(8)}\chi_{R}^{(8)})
                          \bar{\chi}_{NS+}
                        -(\chi_{NS+}^{(8)}\chi_{NS+}^{(8)}
                         +\chi_{R}^{(8)}\chi_{R}^{(8)})
                          \bar{\chi}_{R}\right\}  \nonumber \\
    & & \hspace*{5mm}
 +{\cal Z}_{(1,0)}\left\{2\chi_{R}^{(8)}\chi_{NS-}^{(8)}\bar{\chi}_{NS-}
                        -2\chi_{R}^{(8)}\chi_{NS+}^{(8)}\bar{\chi}_{R}\right\} \nonumber \\
    & & \hspace*{5mm} \left.
 +{\cal Z}_{(1,1)}\left\{2\chi_{R}^{(8)}\chi_{NS+}^{(8)}\bar{\chi}_{NS-}
                        -2\chi_{R}^{(8)}\chi_{NS-}^{(8)}\bar{\chi}_{R}\right\} \right] . 
     \nonumber \\
\end{eqnarray}
In the $R\to 0$ limit, 
\begin{eqnarray}
Z_F &\propto&   (\chi_{NS+}^{(8)}\chi_{NS+}^{(8)}
                         +\chi_{R}^{(8)}\chi_{R}^{(8)})
                          \bar{\chi}_{NS+}
                        -(\chi_{NS-}^{(8)}\chi_{NS-}^{(8)}
                         +\chi_{R}^{(8)}\chi_{R}^{(8)})
                          \bar{\chi}_{R}
                \nonumber \\
   & &         +2\chi_{R}^{(8)}\chi_{NS-}^{(8)}\bar{\chi}_{NS-}
                        -2\chi_{R}^{(8)}\chi_{NS+}^{(8)}\bar{\chi}_{R}.
  \label{SO(16)2}
\end{eqnarray}
This coincides with the fermion part of the partition function for the non-supersymmetric 
$SO(16)\times SO(16)$ theory.

\vspace{5mm}

\subsection{The $E_8\times E_8$ theory}

\vspace{3mm}

(v) $\delta = (1,\ 0^7;\ 0^8)$

\vspace{2mm}

From the similar calculation to the one in the case (i), 
\begin{eqnarray}
Z_F &=& 8(\chi_{NS+}^{(8)}+\chi_R^{(8)})\left\{\ \ 
  {\cal Z}_{(0,0)}(\chi_{NS+}^{(8)}\bar{\chi}_{NS+}-\chi_R^{(8)}\bar{\chi}_R)
 +{\cal Z}_{(0,1)}(\chi_{R}^{(8)}\bar{\chi}_{NS+}-\chi_{NS+}^{(8)}\bar{\chi}_R)
  \right. \nonumber \\
    & & \hspace{3cm} \left.
 +{\cal Z}_{(1,0)}(\chi_{NS-}^{(8)}\bar{\chi}_{NS-}-\chi_R^{(8)}\bar{\chi}_R)
 +{\cal Z}_{(1,1)}(\chi_{R}^{(8)}\bar{\chi}_{NS-}-\chi_{NS-}^{(8)}\bar{\chi}_R)
  \right\}. \nonumber \\
\end{eqnarray}
The fermion part of the partition function for the non-supersymmetric $E_8\times SO(16)$ 
theory is obtained in the $R\to 0$ limit. 
\begin{equation}
Z_F \ \propto\  (\chi_{NS+}^{(8)}+\chi_R^{(8)})
            (\chi_{NS+}^{(8)}\bar{\chi}_{NS+}
            +\chi_{NS-}^{(8)}\bar{\chi}_{NS-}-2\chi_R^{(8)}\bar{\chi}_R)
\end{equation}

\vspace{6mm}

(vi) $\delta = \left(\left(\mbox{$\frac12$}\right)^2,\ 0^6;\ 
                     \left(\mbox{$\frac12$}\right)^2,\ 0^6\right)$

\vspace{2mm}

The partition function is a bit complicated in this case, 
\begin{eqnarray}
Z_F &=& 8\left(\ \ 
  {\cal Z}_{(0,0)}\left[\left\{(\chi_{NS+}^{(2)}\chi_{NS+}^{(6)}+\chi_R^{(2)}\chi_R^{(6)})^2
                              +(\chi_{NS-}^{(2)}\chi_{NS-}^{(6)}+\chi_R^{(2)}\chi_R^{(6)})^2
                        \right\}\bar{\chi}_{NS+} \right.\right. \nonumber \\
    & & \hspace*{2cm} \left.
                        -2(\chi_{NS+}^{(2)}\chi_{NS+}^{(6)}+\chi_R^{(2)}\chi_R^{(6)})
                          (\chi_{NS-}^{(2)}\chi_{NS-}^{(6)}+\chi_R^{(2)}\chi_R^{(6)})
                          \bar{\chi}_R \right] \nonumber \\
    & & \hspace*{5mm}
 +{\cal Z}_{(0,1)}\left[2(\chi_{NS+}^{(2)}\chi_{NS+}^{(6)}+\chi_R^{(2)}\chi_R^{(6)})
                         (\chi_{NS-}^{(2)}\chi_{NS-}^{(6)}+\chi_R^{(2)}\chi_R^{(6)})
                          \bar{\chi}_{NS+} \right. \nonumber \\
    & & \hspace*{2cm} \left.
                    -\left\{(\chi_{NS+}^{(2)}\chi_{NS+}^{(6)}+\chi_R^{(2)}\chi_R^{(6)})^2
                           +(\chi_{NS-}^{(2)}\chi_{NS-}^{(6)}+\chi_R^{(2)}\chi_R^{(6)})^2
                     \right\}\bar{\chi}_R \right] \nonumber \\
    & & \hspace*{5mm}
 +{\cal Z}_{(1,0)}\left[
               \left\{(\chi_{R}^{(2)}\chi_{NS+}^{(6)}+\chi_{NS-}^{(2)}\chi_R^{(6)})^2
                     +(\chi_{NS+}^{(2)}\chi_{R}^{(6)}+\chi_R^{(2)}\chi_{NS-}^{(6)})^2
                        \right\}\bar{\chi}_{NS-} \right. \nonumber \\
    & & \hspace*{2cm} \left.
                        -2(\chi_{R}^{(2)}\chi_{NS+}^{(6)}+\chi_{NS-}^{(2)}\chi_R^{(6)})
                          (\chi_{NS+}^{(2)}\chi_{R}^{(6)}+\chi_R^{(2)}\chi_{NS-}^{(6)})
                          \bar{\chi}_R \right] \nonumber \\
    & & \hspace*{5mm}
 +{\cal Z}_{(1,1)}\left[2(\chi_{R}^{(2)}\chi_{NS+}^{(6)}+\chi_{NS-}^{(2)}\chi_R^{(6)})
                         (\chi_{NS+}^{(2)}\chi_{R}^{(6)}+\chi_R^{(2)}\chi_{NS-}^{(6)})
                          \bar{\chi}_{NS-} \right. \nonumber \\
    & & \hspace*{2cm} \left. \left.
                    -\left\{(\chi_{R}^{(2)}\chi_{NS+}^{(6)}+\chi_{NS-}^{(2)}\chi_R^{(6)})^2
                           +(\chi_{NS+}^{(2)}\chi_{R}^{(6)}+\chi_R^{(2)}\chi_{NS-}^{(6)})^2
                     \right\}\bar{\chi}_R \right] \right). \nonumber \\
\end{eqnarray}
In the $R\to 0$ limit, 
\begin{eqnarray}
Z_F &\propto& \ \ 
              \left\{(\chi_{NS+}^{(2)}\chi_{NS+}^{(6)}+\chi_R^{(2)}\chi_R^{(6)})^2
                    +(\chi_{NS-}^{(2)}\chi_{NS-}^{(6)}+\chi_R^{(2)}\chi_R^{(6)})^2
                        \right\}\bar{\chi}_{NS+} \nonumber \\
    & &            -2(\chi_{NS+}^{(2)}\chi_{NS+}^{(6)}+\chi_R^{(2)}\chi_R^{(6)})
                     (\chi_{NS-}^{(2)}\chi_{NS-}^{(6)}+\chi_R^{(2)}\chi_R^{(6)})
                          \bar{\chi}_R \nonumber \\
    & &      +\left\{(\chi_{R}^{(2)}\chi_{NS+}^{(6)}+\chi_{NS-}^{(2)}\chi_R^{(6)})^2
                    +(\chi_{NS+}^{(2)}\chi_{R}^{(6)}+\chi_R^{(2)}\chi_{NS-}^{(6)})^2
                        \right\}\bar{\chi}_{NS+}  \nonumber \\
    & &            -2(\chi_{R}^{(2)}\chi_{NS+}^{(6)}+\chi_{NS-}^{(2)}\chi_R^{(6)})
                     (\chi_{NS+}^{(2)}\chi_{R}^{(6)}+\chi_R^{(2)}\chi_{NS-}^{(6)})
                          \bar{\chi}_R .
\end{eqnarray}
This is the fermion part of the partition function for the non-supersymmetric 
$(E_7\times SU(2))^2$ theory. 
Note that $\gamma_\delta$ can be divided as $\gamma_1\gamma_2$, where 
\begin{equation}
\gamma_{1,2} = (-1)^\epsilon(-1)^{F_{1,2}^{(2)}}. 
\end{equation}
The operator $F_1^{(2)}$ counts the number of the $\lambda_L^{1,2}$ and the operator 
$F_2^{(2)}$ the number of the $\lambda_L^{9,10}$, and 
\begin{equation}
\epsilon = \left\{
  \begin{array}{ll}
     0, & \mbox{for NS-sector,} \\ 1, & \mbox{for R-sector.}
  \end{array}
\right.
\end{equation}

\vspace{6mm}

(vii) $\delta = (1,\ 0^7;\ 1,\ 0^7)$

\vspace{2mm}

One can obtain the following. 
\begin{eqnarray}
Z_F &=& 8\left[\ \ 
  {\cal Z}_{(0,0)}\left\{(\chi_{NS+}^{(8)}\chi_{NS+}^{(8)}+\chi_R^{(8)}\chi_R^{(8)})
                         \bar{\chi}_{NS+}
                        -2\chi_{NS+}^{(8)}\chi_R^{(8)}\bar{\chi}_R\right\}\right.
     \nonumber \\
    & & \hspace*{5mm}
 +{\cal Z}_{(0,1)}\left\{2\chi_{NS+}^{(8)}\chi_R^{(8)}\bar{\chi}_{NS+}
                        -(\chi_{NS+}^{(8)}\chi_{NS+}^{(8)}+\chi_R^{(8)}\chi_R^{(8)})
                         \bar{\chi}_{R}\right\} \nonumber \\
    & & \hspace*{5mm}
 +{\cal Z}_{(1,0)}\left\{2\chi_{NS-}^{(8)}\chi_R^{(8)}\bar{\chi}_{NS-}
                        -(\chi_{NS-}^{(8)}\chi_{NS-}^{(8)}+\chi_R^{(8)}\chi_R^{(8)})
                         \bar{\chi}_{R}\right\} \nonumber \\
    & & \hspace*{5mm} \left.
 +{\cal Z}_{(1,1)}\left\{(\chi_{NS-}^{(8)}\chi_{NS-}^{(8)}+\chi_R^{(8)}\chi_R^{(8)})
                         \bar{\chi}_{NS-}
                        -2\chi_{NS-}^{(8)}\chi_R^{(8)}\bar{\chi}_R\right\}\right],
     \nonumber \\
\end{eqnarray}
and, in the $R\to 0$ limit, 
\begin{eqnarray}
Z_F &\propto& (\chi_{NS+}^{(8)}\chi_{NS+}^{(8)}+\chi_R^{(8)}\chi_R^{(8)})
                         \bar{\chi}_{NS+}
                        -2\chi_{NS+}^{(8)}\chi_R^{(8)}\bar{\chi}_R \nonumber \\
    & &+2\chi_{NS-}^{(8)}\chi_R^{(8)}\bar{\chi}_{NS-}
                        -(\chi_{NS-}^{(8)}\chi_{NS-}^{(8)}+\chi_R^{(8)}\chi_R^{(8)})
                         \bar{\chi}_{R}.
\end{eqnarray}
The last expression is the same as (\ref{SO(16)2}), although the expressions before taking 
the limit are different from each other.


\newpage

\begin{thebibliography}{99}

\bibitem{Matrix}
T.Banks, W.Fischler, S.H.Shenker, L.Susskind, 
{\it M Theory As A Matrix Model: A Conjecture}, 
Phys. Rev. {\bf D55} (1997) 5112, hep-th/9610043; 

N.Ishibashi, H.Kawai, Y.Kitazawa, A.Tsuchiya, 
{\it A Large-N Reduced Model as Superstring}, 
Nucl. Phys. {\bf B498} (1997) 467, hep-th/9612115.

\bibitem{Sen}
For a review, see e.g. \ 
A.Sen, 
{\it Non-BPS States and Branes in String Theory}, 
hep-th/9904207.

\bibitem{WittenSFT}
E.Witten, {\it Non-commutative Geometry and String Field Theory}, 
Nucl. Phys. {\bf B268} (1986) 253. 

\bibitem{OSFT}

L.Rastelli, A.Sen, B.Zwiebach, 
{\it String Field Theory Around the Tachyon Vacuum}, 
hep-th/0012251; 
{\it Classical Solutions in String Field Theory Around the Tachyon Vacuum}, 
hep-th/0102112; 
{\it Half-strings, Projectors, and Multiple D-branes in Vacuum String Field Theory}, 
hep-th/0105058, 

I.Ellwood, W.Taylor, 
{\it Open String Field Theory Without Open Strings}, 
Phys. Lett. {\bf B512} (2001) 181, hep-th/0103085;

D.J.Gross, W.Taylor, 
{\it Split String Field Theory I}, 
JHEP {\bf 0108} (2001) 009, hep-th/0105059; \ 
{\it Split String Field Theory II}, 
JHEP {\bf 0108} (2001) 010, hep-th/0106036. 

T.Kawano, K.Okuyama, 
{\it Open String Fields As Matrices}, 
JHEP {\bf 0106} (2001) 061, hep-th/0105129. 

\bibitem{CostaGutperle}
M.S.Costa, M.Gutperle, 
{\it The Kaluza-Klein Melvin Solution in M-theory}, 
JHEP {\bf 0103} (2001) 027, hep-th/0012072. 

\bibitem{fluxbrane}
M.Gutperle, A.Strominger, 
{\it Fluxbranes in String Theory}, 
JHEP {\bf 0106} (2001) 035, hep-th/0104136. 

\bibitem{RT}
J.G.Russo, A.A.Tseytlin, 
{\it Magnetic Backgrounds and Tachyonic Instabilities in Closed 
Superstring Theory and M-theory}, 
Nucl. Phys. {\bf B611} (2001) 93, hep-th/0104238.

\bibitem{Melvin}
M.A.Melvin, 
{\it Pure Magnetic and Electric Geons}, 
Phys. Lett. {\bf B8} (1964) 65. 

\bibitem{Suyama}
T.Suyama, 
{\it Closed String Tachyons in Non-supersymmetric Heterotic Theories}, 
JHEP {\bf 0108} (2001) 037, hep-th/0106079. 

\bibitem{nonSUSYhet}
L.J.Dixon, J.A.Harvey, 
{\it String Theories in Ten Dimensions Without Spacetime Supersymmetry}, 
Nucl. Phys. {\bf B274} (1986) 93.

\bibitem{nonSUSYhet2}

L.Alvarez-Gaume, P.Ginsparg, G.Moore, C.Vafa, 
{\it An $O(16)\times O(16)$ Heterotic String}, 
Phys. Lett. {\bf B171} (1986) 155; 

N.Seiberg, E.Witten, 
{\it Spin Structures in String Theory}, 
Nucl. Phys. {\bf B276} (1986) 272.

\bibitem{list}
H.Kawai, D.C.Lewellen, S.-H.H.Tye, 
{\it Classification of Closed-fermionic-string Models}, 
Phys. Rev. {\bf D34} (1986) 3794.

\bibitem{fluxbrane2}
P.M.Saffin, 
{\it Gravitating Fluxbranes}, 
Phys.Rev. {\bf D64} (2001) 024014, gr-qc/0104014; 

M.S.Costa, C.A.R.Herdeiro, L.Cornalba, 
{\it Flux-branes and the Dielectric Effect in String Theory}, 
hep-th/0105023. 

\bibitem{motl}
L.Motl, 
{\it Melvin Matrix Models}, 
hep-th/0107002. 

\bibitem{fluxbrane3}
R.Emparan, 
{\it Tubular Branes in Fluxbranes}, 
Nucl. Phys. {\bf B610} (2001) 169, hep-th/0105062. 

\bibitem{fluxbrane4}
D.Brecher, P.M.Saffin, 
{\it A Note on the Supergravity Description of Dielectric Branes}, 
Nucl.Phys. {\bf B613} (2001) 218, hep-th/0106206. 

\bibitem{instability}
F.Dowker, J.P.Gauntlett, D.A.Kastor, J.Traschen, 
{\it Pair Creation of Dilaton Black Holes}, 
Phys. Rev. {\bf D49} (1994) 2909, hep-th/9309075; 

F.Dowker, J.P.Gauntlett, S.B.Giddings, G.T.Horowitz, 
{\it On Pair Creation of Extremal Black Holes and Kaluza-Klein Monopoles}, 
Phys. Rev. {\bf D50} (1994) 2662, hep-th/9312172; \ 
{\it The Decay of Magnetic Fields in Kaluza-Klein Theory}, 
Phys. Rev {\bf D52} (1995) 6929, hep-th/9507143; \ 
{\it Nucleation of P-Branes and Fundamental Strings}, 
Phys. Rev. {\bf D53} (1996) 7115, hep-th/9512154. 

\bibitem{instanton}
R.C.Myers, M.J.Perry, 
{\it Black Holes in Higher Dimensional Space-times}, 
Ann. Phys. {\bf 172} (1986) 304.

\bibitem{bubble}
E.Witten, 
{\it Instability of the Kaluza-Klein Vacuum}, 
Nucl. Phys. {\bf B195} (1982) 481. 

\bibitem{Type0}
O.Bergman, M.R.Gaberdiel, 
{\it Dualities of Type 0 Strings}, 
JHEP {\bf 9907} (1999) 022, hep-th/9906055. 

\bibitem{RussoTseytlin}
A.A.Tseytlin, 
{\it Closed Superstrings in Magnetic Field: Instabilities and
Supersymmetry Breaking}, 
Nucl. Phys. Proc. Suppl. {\bf 49} (1996) 338, hep-th/9510041; 

J.G.Russo, A.A.Tseytlin, 
{\it Magnetic Flux Tube Models in Superstring Theory}, 
Nucl. Phys. {\bf B461} (1996) 131, hep-th/9508068. 

\bibitem{conformal}
G.T.Horowitz, A.A.Tseytlin, 
{\it A New Class of Exact Solutions in String Theory}, 
Phys. Rev. {\bf D51} (1995) 2896, hep-th/9409021; 

J.G.Russo, A.A.Tseytlin, 
{\it Constant Magnetic Field in Closed String Theory: An Exactly Solvable Model}, 
Nucl. Phys. {\bf B448} (1995) 293, hep-th/9411099. 

\bibitem{CMP}
L.Alvarez-Gaume, G.Moore, C.Vafa, 
{\it Theta Functions, Modular Invariance, and Strings}, 
Comm. Math. Phys. {\bf 106} (1986) 1.

\bibitem{toroidal}
K.S.Narain, M.H.Sarmadi, E.Witten, 
{\it A Note on Toroidal Compactification of Heterotic String Theory}, 
Nucl. Phys. {\bf B279} (1987) 369. 

\bibitem{T-duality}
M.Rocek, E.Verlinde, 
{\it Duality, Quotients, and Currents}, 
Nucl. Phys. {\bf B373} (1992) 630, hep-th/9110053; 

E.Bergshoeff, I.Entrop, R.Kallosh, 
{\it Exact Duality in String Effective Action}, 
Phys. Rev. {\bf D49} (1994) 6663, hep-th/9401025; 

E.Alvarez, L.Alvarez-Gaume, I.Bakas, 
{\it T-duality and Spacetime Supersymmetry}, 
Nucl. Phys. {\bf B457} (1995) 3, hep-th/9507112. 

\bibitem{0dual}
O.Bergman, M.R.Gaberdiel, 
{\it A Non-Supersymmetric Open String Theory and S-Duality}, 
Nucl. Phys. {\bf B499} (1997) 183, hep-th/9701137; 

Y.Michishita, 
{\it D0-branes in SO(32)$\times$SO(32) Open Type 0 String Theory}, 
Phys.Lett. {\bf B466} (1999) 161, hep-th/9907094.

\bibitem{Blum}
J.D.Blum, K.R.Dienes, 
{\it Duality without Supersymmetry: The Case of the SO(16) $\times$ SO(16) String}, 
Phys. Lett. {\bf B414} (1997) 260, hep-th/9707148; \ 
{\it Strong/Weak Coupling Duality Relations for Non-Supersymmetric String Theories}, 
Nucl. Phys. {\bf B516} (1998) 83, hep-th/9707160. 

\bibitem{PolWitten}
J,Polchinski, E.Witten, 
{\it Evidence for Heterotic - Type I String Duality}, 
Nucl. Phys. {\bf B460} (1996) 525, hep-th/9510169. 

\bibitem{Sagnotti}
M.Bianchi, A.Sagnotti, 
{\it On the Systematics of Open String Theories}, 
Phys. Lett. {\bf B247} (1990) 517; 

A.Sagnotti, 
{\it Some Properties of Open-String Theories}, 
hep-th/9509080. 

\bibitem{CHL}
S.Chaudhuri, G.Hockney, J.D.Lykken, 
{\it Maximally Supersymmetric String Theories in D$<$10}, 
Phys. Rev. Lett. {\bf 75} (1995) 2264, hep-th/9505054; 

S.Chaudhuri, J.Polchinski, 
{\it Moduli Space of CHL Strings}, 
Phys. Rev. {\bf D52} (1995) 7168, hep-th/9506048.

\bibitem{WilsonLine}
H.M.S.Coxeter, 
{\it Integral caylay numbers}, in Twelve geometric essays, 
Southern Illinois University Press. 




\end{thebibliography}
\end{document}